\documentclass[sigconf,screen,nonacm]{acmart}

\setcopyright{none}

\renewcommand\footnotetextcopyrightpermission[1]{}

\newcommand{\tool}{\textit{LLMEI}}

\begin{document}

\title{Understanding LLM Hallucinations Caused by Soft Errors}

%


\author{Jinghao Wen}
\email{jwen01@villanova.edu}
\affiliation{%
  \institution{Villanova University}
  \department{Department of Electrical and Computer Engineering}
  \city{Villanova}
  \state{Pennsylvania}
  \postcode{19085}
  \country{USA}
}

\author{Dongning Ma}
\email{Dongning.Ma@mbzuai.ac.ae}
\affiliation{%
  \institution{Mohamed bin Zayed University of Artificial Intelligence}
  \department{Department of Computer Science}
  \city{Abu Dhabi}
  \country{United Arab Emirates}
}

\author{Ruixuan Wang}
\email{rwang8@villanova.edu}
\affiliation{%
  \institution{Villanova University}
  \department{Department of Electrical and Computer Engineering}
  \city{Villanova}
  \state{Pennsylvania}
  \postcode{19085}
  \country{USA}
}

\author{Shaohuang Wang}
\email{swang08@villanova.edu}
\affiliation{%
  \institution{Villanova University}
  \department{Department of Electrical and Computer Engineering}
  \city{Villanova}
  \state{Pennsylvania}
  \postcode{19085}
  \country{USA}
}

\author{Xun Jiao}
\email{xun.jiao@villanova.edu}
\affiliation{%
  \institution{Villanova University}
  \department{Department of Electrical and Computer Engineering}
  \city{Villanova}
  \state{Pennsylvania}
  \postcode{19085}
  \country{USA}
}

\renewcommand{\shortauthors}{Wen et al.}

\begin{abstract}
  While hallucinations in Large Language Models (LLMs) have been extensively studied from a software perspective, limited attention has been paid to hardware-level factors. In this paper, we performed a case study to understand LLM hallucination caused by soft errors on hardware. We introduce \tool{} to inject soft errors into LLMs during inference, based on what we evaluate the impact of soft error across diverse LLM architectures and tasks, including mathematical reasoning and code generation. We summarize the taxonomy of different failure modes based on LLM hallucination behaviors and analyze their distributions under different error rates. Our findings underscore the hallucination-specific diagnostics within the reliability of LLM implementation on hardware.
\end{abstract}



\received{1 April 2026}

\maketitle

\section{Introduction}
Large Language Models (LLMs) are rapidly reshaping the AI landscape, with models like GPT-4~\cite{achiam2023gpt}, LLaMA~\cite{touvron2023llama}, Gemini~\cite{team2023gemini}, and DeepSeek~\cite{guo2025deepseek} demonstrating capabilities that surpass human performance in various tasks. Initially designed for conversational AI, LLMs have evolved to exhibit advanced reasoning and logical abilities, expanding their applications across domains such as code generation~\cite{roziere2023code}, scientific research, and autonomous agents~\cite{wang2024survey}.

However, LLMs face critical reliability challenges. The enormous scaling increases the vulnerability of LLMs against soft errors, because of the growing complexity and heterogeneity of modern data infrastructures~\cite{baumann2005soft, dixit2011impact, shivakumar2002modeling}. Soft errors, also known as transient faults, are non-permanent hardware-induced errors that do not cause physical damage to the device but can still corrupt data during execution. Soft errors typically manifest as random bit flips in model parameters or intermediate computations, often stemming from hardware defects or environmental factors~\cite{chang2018evaluating, stutz2021bit, sullivan2021characterizing}. These errors are transient and may disappear after re-execution, making them difficult to detect and debug, especially in large-scale distributed systems such as those used for LLM training and inference. 

Another major concern is the hallucination, which refers to the misalignment of LLM generations from the world knowledge or input prompts~\cite{zhang2025law,ravichander2025halogen,bang2025hallulens}. In particular, HALOGEN~\cite{ravichander2025halogen} summarizes the error categories based on the source of LLM hallucinations. Although hallucinations in LLMs have been widely investigated from the software perspective, such as prompting, decoding choices, or data biases, much less is known about soft errors stemming from hardware that contribute to such failures \cite{bang2025hallulens, huang2025survey}. 



\begin{figure*}
    \centering
    \includegraphics[width=\linewidth]{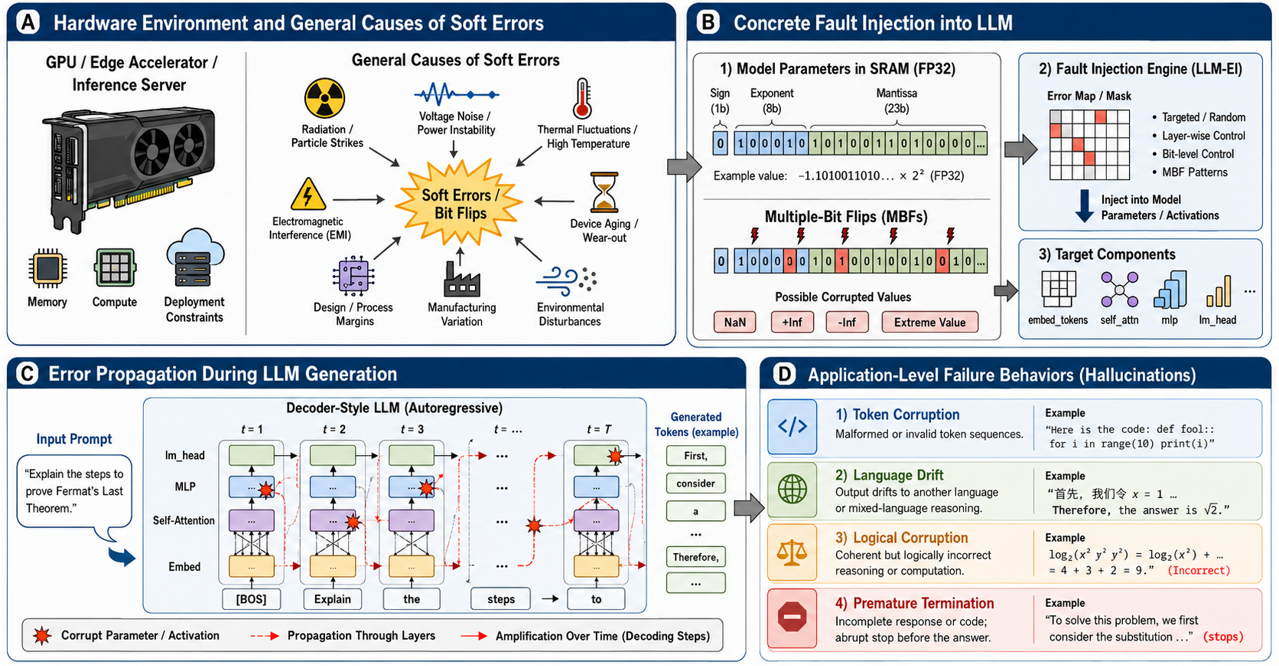}
    \caption{The hallucination of LLMs after efficient and scalable bit-level error injection through \tool{}.}
    \label{fig:corrupted}
\end{figure*}

In this paper, we performed the first case study that systematically investigates how different scales of soft errors can affect the hallucination behavior of the generation of LLMs, as indicated in Figure~\ref{fig:corrupted}. We introduce the first taxonomy of hallucination triggered by soft errors at the application level. We develop a highly efficient, flexible, and user-friendly error injection framework \tool{}, for the soft error evaluation on LLMs. We have open-sourced \tool{}, and the complete code and documentation are included in the supplementary materials.

The major contributions of this paper are as follows:
\begin{itemize}

    \item For the first time, we investigate and report that soft errors can trigger hallucinations during LLM generation. Specifically, we summarize the taxonomy of soft-error-induced LLM hallucination, identifying four main hallucination patterns. 

    \item Additionally, we comprehensively analyze the distribution of different hallucination types across varying error rates, model architectures, tasks, and internal components. This analysis characterizes how hardware-induced soft errors manifest as distinct application-level generation failures and how their distributions evolve as fault severity increases.

    \item To achieve large-scale LLM error injection simulations, we develop and open-source \tool{}, which is specifically designed to enable large-scale error injection campaigns to LLMs efficiently and flexibly. In particular, \tool{} supports highly efficient and flexible error injection simulation on LLMs with different scales.


\end{itemize}

\section{Related Works}
Hallucination is a major reliability challenge in LLMs. Existing studies have shown that LLMs can generate fluent but incorrect, unfaithful, or unsupported content, which severely limits their trustworthiness in high-stakes applications such as mathematical reasoning, code generation, scientific assistance, and autonomous agents~\cite{ji2023survey, bang2025hallulens}. Prior work on hallucination has mainly focused on software-level causes, including insufficient or biased training data, imperfect alignment, decoding strategies, retrieval failures, and prompt sensitivity~\cite{maynez2020faithfulness, zhang2025siren}. Accordingly, existing mitigation methods are typically designed from the algorithmic perspective, such as retrieval-augmented generation, uncertainty estimation, self-consistency, post-hoc verification, and hallucination detection~\cite{manakul2023selfcheckgpt, ke2024critiquellm}. 

However, there is very limited study on if soft errors on hardware can directly trigger hallucination-like behaviors for LLMs. Soft errors are typically caused by transient hardware disturbances such as radiation-induced particle strikes, voltage noise, or device aging, which may lead to bit flips in memory or computation units~\cite{baumann2005soft, shivakumar2002modeling}. With the rapid growth of model size and deployment scale, particularly in large neural networks, the impact of such low-level numerical faults can propagate through model computations and lead to unpredictable behaviors in downstream applications.
This gap motivates our study of soft-error-induced hallucinations, where transient bit-level faults in model parameters can manifest as high-level generation failures.

\begin{figure*}[htbp]
    \centering
    \includegraphics[width=\linewidth]{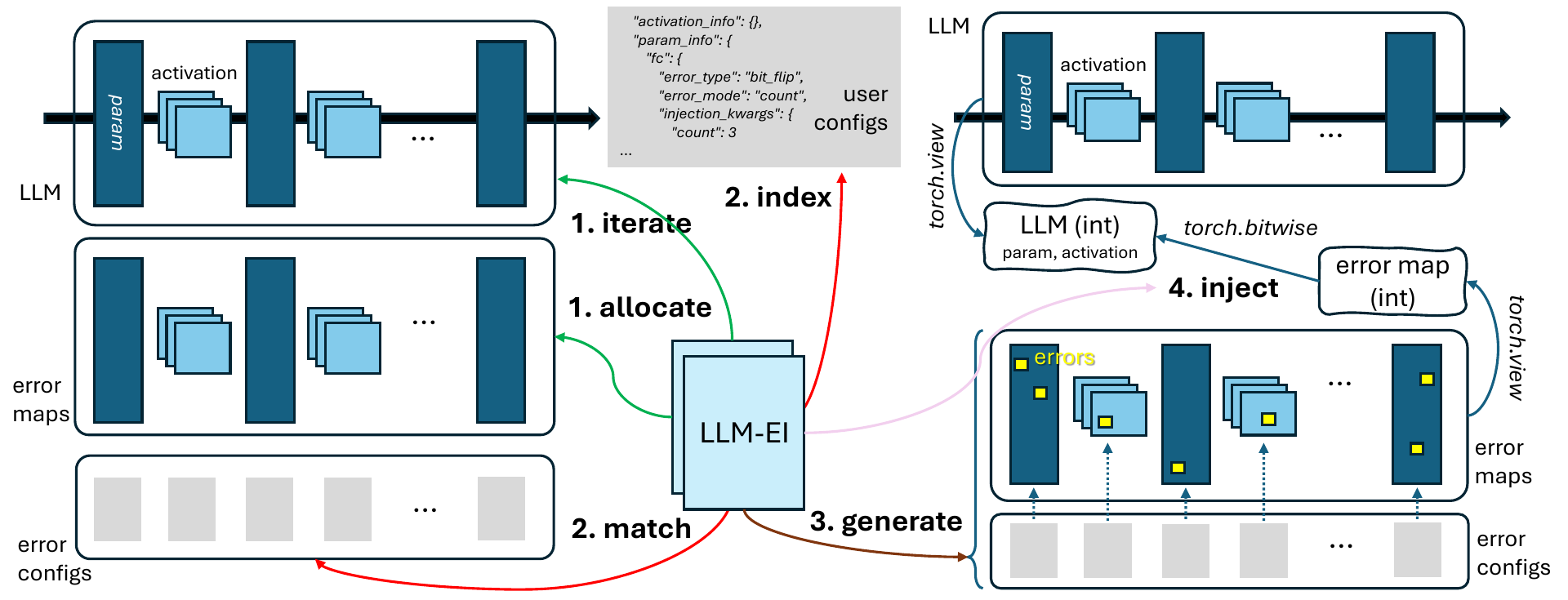}
    \caption{Overview of \tool.}
    \label{fig:framework}
\end{figure*}

\section{Error Injection}
\subsection{Background and Scope}

We consider the deployment of LLMs on modern hardware platforms such as GPUs, edge accelerators, and large-scale inference servers, where computational reliability is constrained. In practical systems, aggressive optimization techniques, such as voltage scaling, reduced-precision computation, and limited error protection, are commonly adopted to improve energy efficiency and throughput. However, these operating conditions inevitably increase the susceptibility of hardware components to soft errors, including radiation-induced bit flips, voltage noise, and device aging effects. Prior studies have shown that lowering supply voltage can significantly increase the bit error rate (BER) in on-chip memories and arithmetic units, making model parameters and intermediate activations more vulnerable to corruption. When these corrupted values propagate through deep neural architectures, they may trigger large numerical perturbations and destabilize the inference process, leading to severe performance degradation~\cite{mahmoud2022goldeneye, reagen2018ares}.

For LLMs, the impact of such faults can be especially pronounced. Bit flips in floating-point representations may produce abnormal numerical values such as NaN or Inf, which propagate through attention mechanisms and feed-forward layers, disrupting the stability of internal representations. Moreover, due to the autoregressive nature of LLM generation, an error occurring at a single step can influence subsequent token predictions and progressively amplify over the generation process. As a result, soft errors may lead to incoherent outputs, semantic drift, or abrupt generation failures, highlighting the practical importance of studying and mitigating soft-error vulnerabilities in LLM deployment environments. 

In this paper, we target the soft errors manifested as multiple bit-flips in the SRAM on DNN accelerator, characterized by the specific bit error rate, which is consistent with prior research~\cite{reagen2018ares, stutz2021bit} and widely observed in the industrial field. To mimic the real context of soft errors, we simulate the occurrence of multiple-bit flips in memory and conduct error injection in the DNN model parameters.

\subsection{Workflow of \tool}
While there are existing works related to error and fault injection to machine learning models~\cite{reagen2018ares, stutz2021bit}, they cannot be directly applied in our case because LLMs are significantly larger; hence, most of the prior frameworks fail to complete error injection within a feasible time. We develop \tool{}, a fast, flexible and scalable framework for error injection into LLMs. Figure~\ref{fig:framework} shows an overview of using \tool{} to inject bit flip errors into LLMs. It has four major steps: 1. \tool{} \textbf{iterates} through the target LLM for error injection, and \textbf{allocates} empty error maps as tensors based on the targets, such as LLM parameters and activations. 2. For each target parameter, \tool{} \textbf{indexes} the specific error configurations from the user input (a JSON file) and \textbf{matches} them with the allocated error maps. 3. \tool{} \textbf{generates} the error maps based on the matched configurations. 4. \tool{} \textbf{injects} errors to the LLM by applying the error maps. 

\noindent
\textbf{Iterate and Allocate.} Before error injection could happen, \tool{} needs to know more about the target LLM first, e.g., what are the parameters inside, what are the type and shape of the parameters and activations. \tool{} will allocate an error map for each target based on that information. As we focus on bit flip error here, an error map is a binary tensor indicating which bit of the target will be flipped. For example, if a parameter in the LLM is in the shape of $(C, H, W)$ and the data type of it is single precision floating point (\textit{torch.float32}), then the allocated error map will be a binary tensor of $(C, H, W, D)$ where $D$ refers to the bit-width, i.e., $D = 32$. Note that the actual errors will be generated at later stage (Step 3) so right here the error maps are all zeros.

\noindent
\textbf{Index and Match.} Another information \tool{} needs to know is the error configuration. In \tool, user can provide a JSON file indicating the rate of bit flips, or the total number of bit flips to inject for each target. \tool{} will match the error configurations to each parameter for generation.

\noindent
\textbf{Generate.} Thanks to the functionality of PyTorch, generating bit flip errors of a target can be straightforwardly done via \textit{torch.rand} and/or \textit{torch.randperm}. \tool{} has high flexibility, i.e., for each target (parameter and activation), users can choose to generate bit flips based on a probability (bit error rate), or based on a fixed number of flips.

\noindent
\textbf{Inject.} PyTorch unfortunately does not support bit-wise operations between floating point tensors and tensors in other data types. Existing frameworks often rely on third party libraries such as \textit{bitstring} and \textit{struct} which are prohibitively slow. \tool{}, instead of type-casting the tensors, use \textit{torch.view} to directly interpret the tensor into different data types such as \textit{torch.int32}. Since our error maps are also PyTorch tensors, we can fortunately aggregate them into \textit{torch.int32} and perform bit-wise operations using \textit{torch.bitwise}. In this way, we are able to speed-up the overall injection phase by more than 1,000 times.

\section{Experimental Setup}

\textbf{Tasks:} To evaluate the performance of large language models (LLMs) under soft errors, we conduct experiments on two representative benchmarks:

\begin{itemize}
    \item AIME 2024~\cite{aime2024}: A mathematical reasoning dataset designed to test symbolic logic and multi-step computation capabilities in LLMs.
    \item LiveCodeBench~\cite{jain2024livecodebenchholisticcontaminationfree}: A coding benchmark focusing on real-world programming problems, where correctness depends on precise logic and syntax generation.
\end{itemize}

\noindent
\textbf{Models:} We select different LLMs according to the characteristics of the evaluated tasks. For mathematical reasoning on AIME 2024, we evaluate three DeepSeek-R1 distilled models with different backbone architectures and scales: DeepSeek-R1-Distill-Qwen-1.5B, DeepSeek-R1-Distill-Qwen-7B, and DeepSeek-R1-Distill-LLaMA-8B. For code generation on LiveCodeBench, we consider both base and instruction-tuned variants from multiple model families, including DeepSeek-Coder-1.3B and 6.7B, Qwen2.5-Coder-1.5B, 3B, and 7B, as well as LLaMA-3.2-1B, LLaMA-3.2-3B, and LLaMA-3.1-8B~\cite{huggingface}. This task-specific model selection covers different parameter scales, backbone architectures, and specialization strategies, enabling us to examine whether the impact of bit-level soft errors varies across reasoning-oriented and code-generation-oriented LLMs.


\begin{figure}[htbp]  
\centering
\includegraphics[width=\columnwidth]{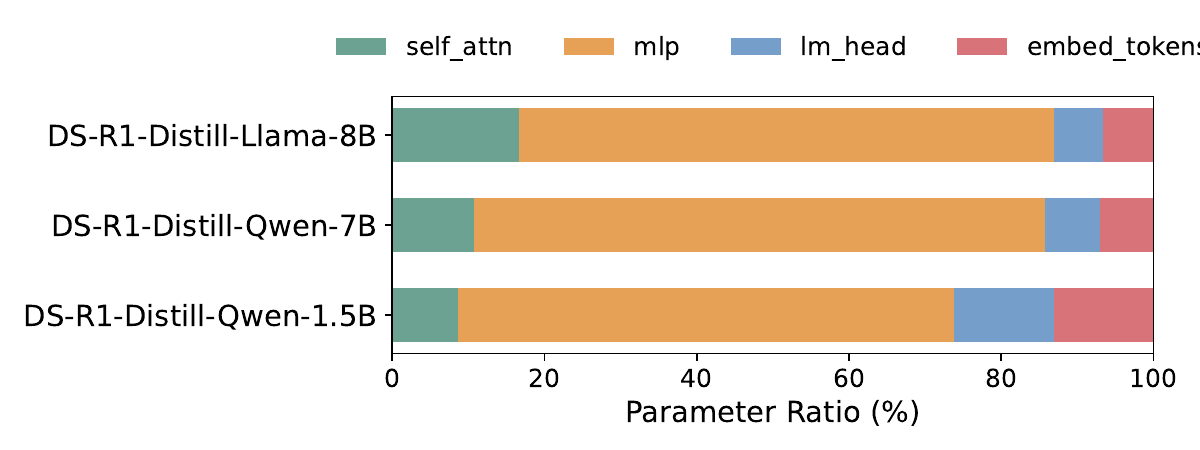}
\caption{Parameter composition of three DS-distill models architectural components.}
\label{fig:distribution}
\end{figure}

As illustrated in Figure~\ref{fig:distribution}, we also analyze the internal parameter composition of three DS-distill models architectural components to guide fine-grained error injection, including \texttt{self\_attn}, \texttt{mlp}, \texttt{lm\_head}, and \texttt{embed\_tokens} layers. 
This component-wise profiling reveals that encoder-related layers (\texttt{mlp} layers) dominate the parameter space, typically accounting for over 60\% of total parameters. \texttt{self\_attn}, \texttt{lm\_head} and \texttt{embed\_tokens} layers comprise a smaller but still significant portion, depending on the model family. 
This decomposition enables targeted error injection into specific components and allows us to observe how localized bit-flip errors propagate into global model behavior. It also facilitates comparative analysis of component robustness across different architectures.

In mathematical reasoning, We evaluate 3 DeepSeek distilled LLMs that differ in architecture and scale: DS-R1-Distilled Qwen-1.5B, Qwen-7B and LLaMA-8B~\cite{guo2025deepseek}. All models are evaluated under identical prompt formats and inference settings across the AIME2024 dataset.

In code generation task, we evaluate state-of-the-art LLMs that span multiple model families and capacities, with their instruction fine-tuned versions, to capture both general and code-specialized LLMs across different architectural designs and parameter scales. All models are tested on the LiveCodeBench benchmark using consistent few-shot prompting strategies and deterministic decoding to ensure fair comparisons. The correctness of generated code is assessed via automated test suites accompanying each task, allowing an objective evaluation of functional correctness under varying error conditions.

\noindent
\textbf{Error injection:} 
We use \tool{} to inject errors into model parameters according to a specified bit error rate (BER) or a fixed number of bit flips, which emulates SRAM-based soft errors commonly observed in real-world hardware platforms~\cite{baumann2005radiation}.

In practical hardware systems, BER can vary significantly depending on device technology, operating voltage, and environmental conditions. As shown in Table~\ref{tab:ber_real_world}, privous studies report that modern ECC-protected DRAM in datacenter environments typically exhibits BERs on the order of $10^{-12}$ to $10^{-11}$, while on-chip SRAM used in DNN accelerators may experience higher error rates around $10^{-10}$ to $10^{-8}$. When aggressive voltage scaling or near-threshold operation is applied to improve energy efficiency, the BER can further increase to approximately $10^{-7}$ or higher. In extreme environments such as radiation-exposed hardware systems or aerospace applications, it may reach $10^{-5}$ or above.

Motivated by these real-world observations, we evaluate model robustness across BER values ranging from $10^{-12}$ to $10^{-4}$, covering both typical deployment conditions and elevated fault regimes observed in energy-efficient or reliability-constrained hardware systems~\cite{reagen2018ares, stutz2021bit}. Under different BER settings, we measure model robustness as the ability to maintain performance under increasing error rates. A rapid performance degradation as BER increases indicates high sensitivity to soft errors.
To reduce statistical variance, each injection configuration is repeated for 10 trials, and the pass@1 score is averaged across runs. For both tasks, we report pass@1 as the primary evaluation metric to measure the functional correctness of generated outputs.

\begin{table}[t]
\centering
\caption{Typical BER ranges observed in real-world hardware. }
\begin{tabular}{lcc}
\hline
Hardware Systems & Typical BER Range \\
\hline
ECC-protected DRAM (Datacenter) & $10^{-12}$ -- $10^{-11}$ ~\cite{schroeder2009dram}\\
On-chip SRAM (Nominal Voltage) & $10^{-10}$ -- $10^{-8}$  \\
SRAM under Voltage Scaling & $10^{-8}$ -- $10^{-7}$  ~\cite{ganapathy2017characterizing}\\
Radiation-Exposed Systems & $10^{-6}$ -- $10^{-5}$  ~\cite{sullivan2021characterizing}\\
\hline
\end{tabular}
\label{tab:ber_real_world}
\end{table}

\noindent
\textbf{Platform:} All experiments are conducted on a server equipped with 4 NVIDIA A800-SXM4-80GB GPUs and 3 NVIDIA RTX A6000-48GB GPUs, alongside dual Intel Xeon Platinum 8358P CPUs. All models and error injection experiments are performed in PyTorch using \tool{}, and we conduct \textbf{3200 GPU hours} evaluation experiments in total.

\section{Results and Analysis}
In this section, we present a comprehensive empirical study on the impact of soft errors across multiple LLMs and tasks.
Under different BERs using \tool{}, we evaluate model resilience, the ability to sustain performance under increasing soft errors, spanning both mathematical reasoning and code generation benchmarks. A rapid performance degradation indicates the high sensitivity of corresponding model to soft errors. The analysis covers model-wise and component-wise error injection across diverse architectures and model scales, as well as the taxonomy and proportion of different LLM hallucination behaviors caused by soft error. Through this, we aim to characterize how different error settings and model structures affect LLM behavior under realistic error conditions.

\subsection{Hallucinations Induced by Soft Errors}

\noindent\fbox{%
\parbox{0.98\linewidth}{%
\textit{\textbf{Observation 1}: Increasing BER induces progressively severe hallucination types, from partial logical errors to total collapse, revealing a staged degradation pattern.}
}}

To gain deeper insight into how bit-level errors propagate into LLM behavior, we quantify the types of hallucination of LLM observed under varying BER conditions. We also categorize corrupted answers of LLM into four distinct types:

\begin{itemize}
    \item \textbf{Token Corruption (Category I):} Outputs that contain nonsensical or non-linguistic token sequences (e.g., repeated or malformed tokens).
    \item \textbf{Language Drift (Category II):} Grammatically correct but linguistically inconsistent outputs, often mixing multiple languages (e.g., English and Chinese).
    \item \textbf{Logical Corruption (Category III):} Grammatically coherent responses with logical or semantic flaws such as contradictory reasoning, incorrect derivations, or loops.
    \item \textbf{Premature Termination (Category IV):} Outputs where the model terminates prematurely without completing the reasoning process.
\end{itemize}

Figure~\ref{fig:corrupted} illustrates representative examples of these four failure modes. Each category highlights a distinct behavioral pattern in corrupted LLM outputs due to bit-level errors. 
These failure types reflect the multifaceted impact of low-level hardware errors on high-level LLM performance, demonstrating that even subtle soft errors can manifest in diverse and unpredictable ways. This categorization provides a foundation for future work in developing targeted error detection and mitigation strategies tailored to each failure mode.

\subsection{Hallucinations Distribution across Tasks}

\noindent\fbox{%
\parbox{0.98\linewidth}{%
\noindent\textit{\textbf{Observation 2}: Different BERs induce distinct hallucination distributions, and LLM robustness varies not only by model size but also by task type.}
}}

\begin{figure}
    \centering
    \includegraphics[width=\linewidth]{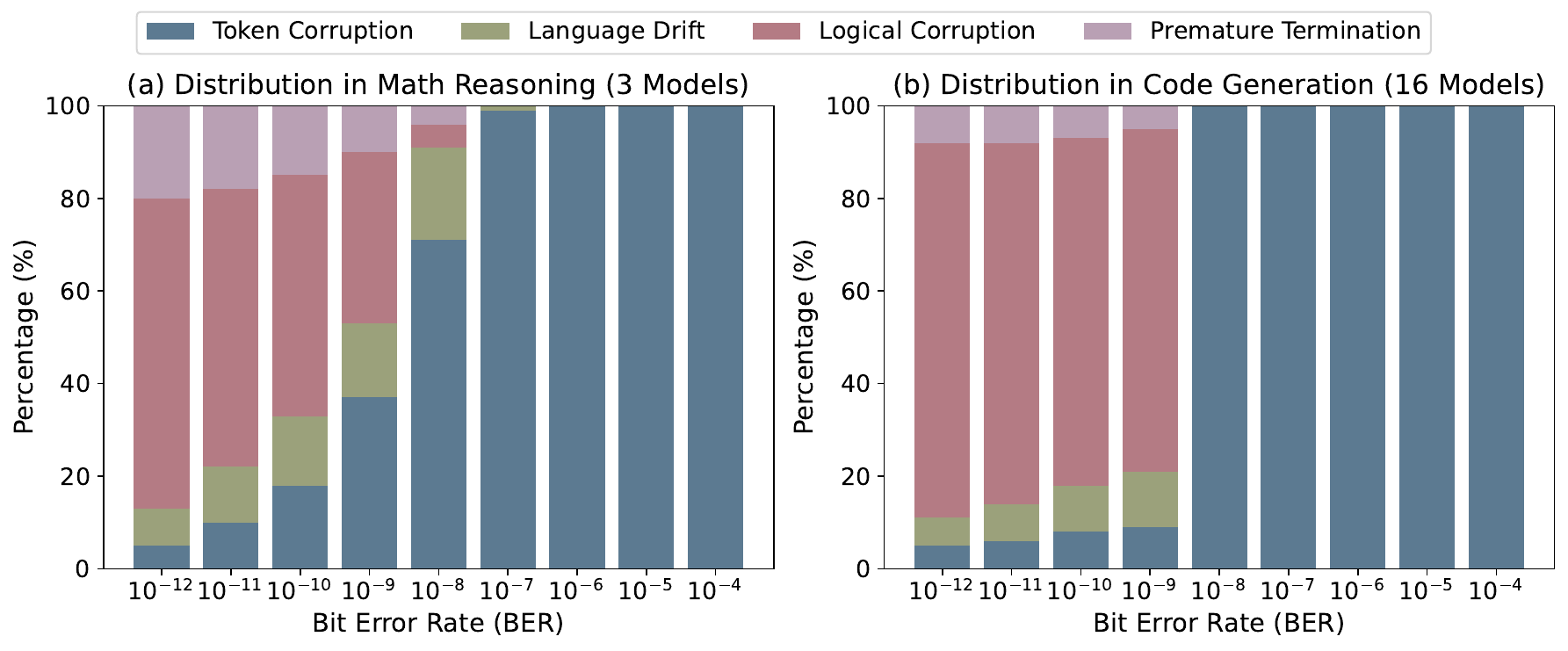}
    \caption{Distribution of corrupted hallucination of different LLMs under varying BER.}
    \label{fig:math bar}
\end{figure}

The progressive distribution transition is clearly visible in Figure~\ref{fig:math bar}(a), where different hallucination types emerge gradually as BER increases. At very low BERs, the dominant failure mode is \textbf{Category III}, indicating that small numbers of bit-level errors mainly disrupt the model’s ability to carry out coherent reasoning, while its language structure remains intact. As BER increases to intermediate levels, we begin to observe the appearance of \textbf{Category II} and \textbf{Category IV}, reflecting growing instability in both semantic consistency and output completeness. Eventually, \textbf{Category I} becomes dominant at high BERs, where the model collapses into producing malformed tokens with little to no structure. This layered degradation pattern highlights how models under mathematical reasoning tasks retain some structural resilience under mild corruption before finally collapsing as errors accumulate.

In contrast, Figure~\ref{fig:math bar}(b) exhibits a much sharper distribution change in the code generation task. The distribution remains largely unchanged up to a BER of $10^{-9}$, with \textbf{Category III} dominating. However, beyond this threshold, \textbf{Category I} abruptly takes over and occupies 100\% of corrupted outputs. This steep collapse reflects the nature of code generation tasks, where outputs must adhere to strict syntactic and semantic constraints to be considered valid. Even minor disruptions in logical flow or structure can render entire generations unusable. In contrast, tasks like mathematical reasoning allow for more flexible interpretations and partial credit, achieving a stronger robustness to soft errors. Overall, Figure~\ref{fig:math_bar} reveals a clear trend:

\begin{itemize}
    \item Small bit errors → primarily Logical Corruption (logical / semantic corruption).
    \item Large bit errors → collapse into Token Corruption (token-level corruption).
\end{itemize}

As a result, different BERs lead to distinct distributions of hallucination types, which underscores the importance of analyzing not just whether a model fails, but how it fails under increasing soft errors. Furthermore, LLM robustness is not only model-dependent, but also task-dependent, error robustness thresholds vary significantly across task types based on their structural and semantic requirements.

\subsection{Hallucinations Distribution across Components}

\noindent\fbox{%
\parbox{0.98\linewidth}{%
\noindent\textit{\textbf{Observation 3}: Different LLM components exhibit distinct hallucination distributions for error, highlighting the importance of component-aware soft errors mitigation.}
}}

\begin{figure}
    \centering
    \includegraphics[width=\linewidth]{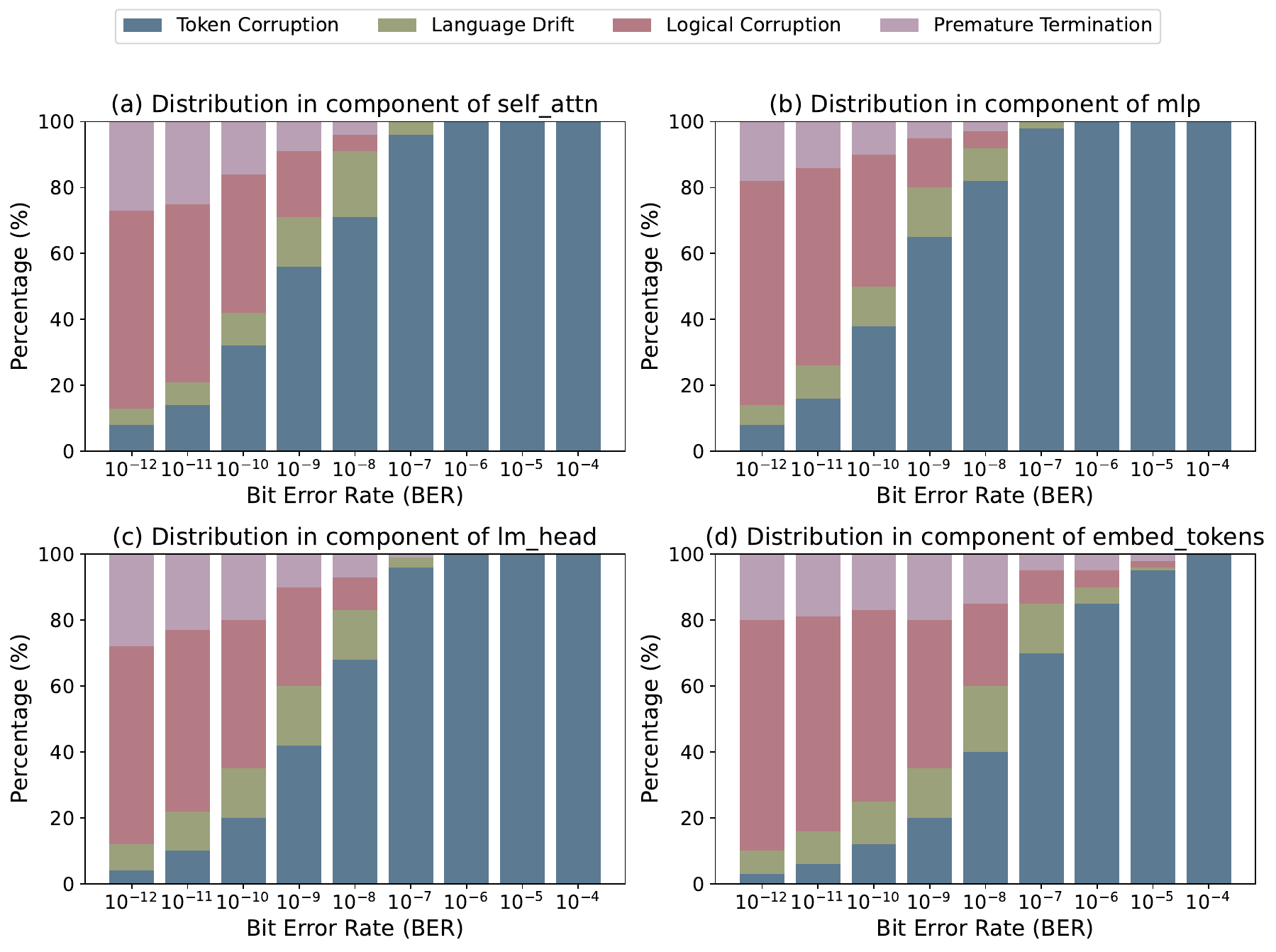}
    \caption{Distribution of corrupted answer of different LLMs and specific components under varying BER.}
    \label{fig:math_bar}
\end{figure}

To further understand how different components of the model contribute to specific failure behaviors, we analyze the average corrupted answer distributions across 4 major components in mathematical reasoning task, as shown in Figure~\ref{fig:math_bar}. The results reveal that \texttt{self\_attn} and \texttt{mlp} are the most sensitive, with rapid transitions from \textbf{Category III} to \textbf{Category I} at moderate BERs ($10^{-10}$-$10^{-8}$), reflecting their central role in maintaining internal reasoning and sequence consistency. \texttt{lm\_head} shows a smoother degradation, exhibiting a mix of \textbf{Category II} and \textbf{Category IV} under BER with $10^{-7}$. In contrast, \texttt{embed\_tokens} remains relatively robust, retaining meaningful behavior up to higher BERs. Despite under high BER ($10^{-7}$), its corruption results in more diverse and less destructive failure types (e.g., \textbf{Category III} or \textbf{Category II}) rather than immediate collapse into \textbf{Category I}. This robustness may stem from the static nature of embedding matrices post-training and the one-time role they play in encoding the input tokens.

\subsection{Model- and Component-Wise Error Robustness}

We additionally examine the soft-error robustness of different model scales and architectural components. The results indicate that robustness varies across models: some larger models begin to experience performance degradation at relatively low BERs, whereas several smaller models maintain non-zero performance over a wider BER range. Nevertheless, model size alone is insufficient to determine robustness, since baseline task capability, model architecture, and parameter organization may also affect the observed degradation.

\begin{figure*}
    \centering
    \includegraphics[width=1\linewidth]{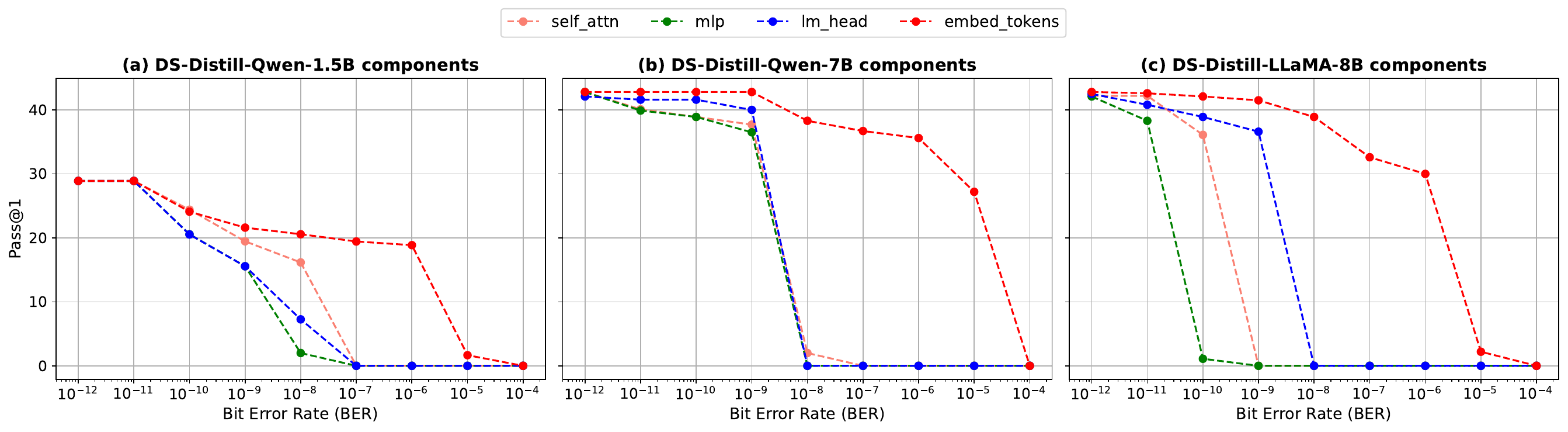}
    \caption{Robustness of different components of 3 DeepSeek-distilled models under different BERs.}
    \label{fig:ber_components}
\end{figure*}

Figure~\ref{fig:ber_components} presents the component-wise robustness of three DeepSeek-R1 distilled models under different BERs. Across the evaluated models, errors injected into \texttt{self\_attn} and \texttt{mlp} generally cause earlier and sharper reductions in Pass@1, indicating that the main Transformer computation blocks are comparatively sensitive to bit-level perturbations. The \texttt{lm\_head} exhibits intermediate sensitivity, while \texttt{embed\_tokens} consistently retains non-zero performance over a substantially wider BER range. These results provide supplementary evidence of heterogeneous error sensitivity across LLM components, while the primary focus of this work remains understanding how hardware-induced soft errors manifest as hallucinations in generated outputs.

\section{Conclusion}

In this paper, we performed a case study to investigate how soft errors in hardware can induce hallucinations in LLMs. Through extensive experiments, we demonstrate that soft errors can propagate through model computation and eventually manifest as high-level hallucinations. We categorize hallucinations into four types: Token Corruption, Language Drift, Logical Corruption, and Premature Termination. Our results further show that the distribution of these hallucinations changes with the bit error rate: relatively mild errors are more likely to disrupt reasoning and semantic consistency, whereas severe corruption can lead to malformed tokens and complete generation collapse.


\bibliographystyle{ACM-Reference-Format}
\bibliography{aaai2026}

\end{document}